\documentclass[a4paper,12pt]{amsart}

\usepackage{amssymb,amsmath,amsthm,graphics}
\theoremstyle{plain}
\newtheorem{theorem}{Theorem}[section]

\newtheorem{lemma}[theorem]{Lemma}

\theoremstyle{remark}
\newtheorem*{remark}{Remark}

\newcommand{\newoperator}[2]{\DeclareMathOperator{#1}{#2}}
\newoperator{\rank}{rank}
\newoperator{\Hom}{H}

\newoperator{\vol}{vol}

\date{}

\begin{document}

\title{Geometric Theory of Lattice Vibrations and Specific Heat}

\author{Mikhail Shubin}

\address{Department of Mathematics,
         Northeastern University,
         360 Huntington Avenue, Boston, MA 02115, USA. \\
E-mail: shubin@neu.edu}

\author{Toshikazu Sunada}

\address{Department of Mathematics,
         Meiji University,
         Higashimita 1-1-1, Tama-ku, Kawasaki 214-8571, Japan \\
E-mail: sunada@math.meiji.ac.jp}

\maketitle

\begin{abstract}
We discuss, from a geometric standpoint, the specific heat of a solid. This is 
a classical subject  in solid state physics  which dates back to  
a pioneering work by Einstein (1907)  and its refinement by Debye (1912). 
Using a special quantization of crystal lattices and calculating the asymptotic 
of the integrated density of states at the bottom of the spectrum, we obtain
a rigorous derivation of the classical Debye $T^3$ law on the specific heat
at low temperatures. The idea and method are taken from discrete geometric analysis 
which has been recently developed for the spectral geometry of crystal lattices.   
\end{abstract}

\section{Introduction}

The primary purpose of this note is to discuss some dynamical properties of solids, 
more specifically {\it lattice vibrations} in {\it crystalline} solids, from a geometric 
view point. In particular, we are concerned with a mathematically sound computation 
of the {\it specific heat}, a typical thermodynamic  quantity in solid state physics. 
The main idea is to employ a technique in {\it discrete geometric analysis} developed 
originally for the study of random walks on {\it crystal lattices} (\cite{kot1}, \cite{kot2}), 
and mathematical apparatus such as {\it von Neumann trace} and {\it direct integrals} 
which make the discussion   more transparent than the existing ones. 

Theoretical computation of the specific heat at low temperature had been 
one of the central themes in quantum physics at the beginning of the last century 
(see \cite{mehra} and \cite{born} for the history). The crucial point in the computation 
is to regard a solid as a crystal lattice realized periodically in the space ${\bf R}^3$. 

To explain what crystal lattices mean, let $V$ be the set of constituent atoms in a solid, 
and $\varPhi:V\longrightarrow {\bf R}^3$ be the injective map representing an arrangement 
of atoms in equilibrium positions. Considering elements in $V$ to be vertices, we join two elements 
in $V$ by an (abstract) edge if they (as atoms) are bound by atomic forces, and extend $\varPhi$ to the set of edges as a piecewise linear map. We thus have a graph $X=(V,E)$ realized in ${\bf R}^3$ where $E$ denotes the set of all oriented edges. When $\varPhi(X)$ is invariant under the action of a lattice $L\subset {\bf R}^3$ by translations, the graph $X$ (or its realization $\varPhi(X)$) is said to be a crystal lattice.

The inter-atomic forces allow the vibrations of atoms which involve small 
excursions from the equilibrium positions. We may describe the vibration 
by the (linearized) equation of motion 
$$
\frac{d^2{\bf f}}{dt^2}=D{\bf f},
$$
where $D$ is a certain linear difference operator of the ``second order" on $X$ 
involving masses of atoms and inter-atomic forces in the coefficients, and 
${\bf f}={\bf f}(t,x)$ stands for the displacement from the equilibrium position 
(thus the position of an atom $x$ at time $t$ is $\varPhi(x)+{\bf f}(t,x)$).

Following Planck's idea on ``energy quanta" originally applied to black-body radiation (1900), or according to the quantum mechanics founded by Heisenberg and Schr\"{o}dinger (1925-26), physicists usually go forward as follows. 

(1) Consider the lattice vibration as an infinite-dimensional system of {\it harmonic oscillators}. 

(2) Decompose the system into independent simple harmonic oscillators, and calculate the distribution of vibration frequencies. 

(3)  Apply statistical mechanics to determine the macroscopic equilibrium state (the {\it Gibbs state}) of the quantized lattice vibration, and compute the {\it internal energy} $U=U(T)$ (per  unit cell) where $T$ is the absolute temperature. Then the {\it specific heat} is given by  
$$
C(T)=\frac{\partial U}{\partial T}.
$$
(See any textbook of solid state physics, for instance, \cite{har}, \cite{kos}, or a review paper 
\cite{bla}, for the detail of this procedure in which one may see a daring manner of physicists 
to bring us effectively to the correct result.  An exception is the book 
\cite{born} by Born and Huang, which, in the second half, is written in a strictly deductive style.)

\medskip

This set routine leads us to the expression 
\begin{eqnarray}\label{internal}
 U(T) = {\text const} +\int_0^{\infty}\frac{\hslash \sqrt{\lambda}}{e^{\hslash \sqrt{\lambda}/KT}-1}d\varphi(\lambda), 
\end{eqnarray}
where 
\begin{eqnarray*}
&&h=2\pi \hslash =\text{Planck constant},\\
&& K=\text{Boltzmann constant}, 
\end{eqnarray*}
and $\varphi(\lambda)$ is the (integrated) {\it density of states}  satisfying the normalization condition 
$$
\int_0^{\infty}~d\varphi(\lambda)=3n \quad  (n=\text{the number of atoms in a unit cell}).
$$
The function $\varphi(\lambda)$ is closely related to the distribution of vibration frequencies (actually, $\sqrt{\lambda}/2\pi$ represents the frequency parameter, and hence $\hslash \sqrt{\lambda}$ stands for ``energy quanta"). As is easily seen,  the behavior of the specific heat at low temperature relies heavily on the asymptotic behavior of $\varphi(\lambda)$ around $\lambda=0$. In the early stages of quantum physics, however, physicists had no rigorous methods, in marked contrast to the case of the black-body radiation, to acquire precise information on $\varphi(\lambda)$ through the microscopic structure of a solid so that they were forced to make daring hypothesis on the shape of $\varphi(\lambda)$ (indeed, it was in 1912 when the discrete structure of solids were confirmed by means of the diffraction of $X$-rays).

The first substantial result was established by A. Einstein in 1907 (two years  
after the publication of his three famous papers; \cite{ein}). He adopted a function 
$\varphi(\lambda)$ defined by 
\begin{eqnarray*}
\varphi(\lambda)=\begin{cases}
0 & (\lambda\leq \lambda_0)\\
3n & (\lambda>\lambda_0)
\end{cases}
\end{eqnarray*}
with one characteristic frequency $\sqrt{\lambda_0}/2\pi$, or equivalently
$$
\frac{d\varphi}{d\lambda}=3n\delta(\lambda-\lambda_0)
$$
(i.e. replacing the vibration spectrum by a set of oscillators at a single frequency) to claim 
$$
C(T)=3nK \Big(\frac{\hslash \sqrt{\lambda_0}}{KT}\Big)^2
\frac{e^{\hslash \sqrt{\lambda_0}/KT}}{(e^{\hslash \sqrt{\lambda_0}/KT}-1\big)^2}.
$$
In spite of the seemingly unrealistic model, this formula explains well not only the {\it law of Dulong and Petit} at high temperature :
$$
C(T)\equiv 3nK,
$$
which had been known since 1791 in the setting of classical mechanics, but also the qualitative fact that, if $T$ goes to zero, then so does $C(T)$ as experimental results show (H. Nernst; 1910). Note, however, that Einstein's formula for $C(T)$ has exponential decay as $T\downarrow 0$, which turns out to be quite incorrect.

In 1912 (\cite{debye6}), Debye proposed, without any knowledge of the lattice structure of crystals as in Einstein's case, to take  the function 
\begin{equation}\label{debye1}
\varphi(\lambda)=\begin{cases}
0 & (\lambda\leq 0)\\
c_0\lambda^{3/2} & (0\leq \lambda \leq \lambda_D)\\
c_0\lambda_D^{3/2} & (\lambda \geq \lambda_D)
\end{cases},
\end{equation}
with a suitable positive constant $c_0$. The quantity $\lambda_D$ is taken so as to satisfy   
$$
\int_0^{\lambda_D}d\varphi(\lambda)=3n.
$$

Debye's distribution $\varphi$ may be inferred from the intuitive observation that the {\it continuum limit} of a crystal lattice is a (uniform) elastic body, and that, in the limit, the density of states for lattice vibrations may be replaced by the one for {\it elastic waves} in a region of low frequencies. Actually, the constant $c_0$ is determined in such a way that the function $c_0\lambda^{3/2}$ is the integrated density states for elastic waves (see Section \ref{remark}). This view is natural because to one 
not aware of the atomic constitution of solids, a solid appears as an elastic continuum. In particular, it is deduced that, if the elastic body is {\it isotropic}, then the constant $c_0$ is given by 
$$
c_0=\frac{{\bf V}}{6\pi^2} \Big(\frac{1}{c_l^3}+\frac{2}{c_t^3}\Big),
$$
where
\begin{eqnarray*}
&&c_l=\text{the longitudinal phase velocity},\\
&&c_t=\text{the transverse phase velocity},\\
&&{\bf V}=\text{the volume of the unit cell}
\end{eqnarray*}
(as a matter of fact, the crystalline solids are never isotropic, so that the constant $c_0$  above should be replaced by an average of phase velocities over all directions of propagation; see \cite{born} or Section \ref{section7}).

With his choice of the distribution $\varphi$, Debye gave the following neat formula for the specific heat 
$$
C(T)=9nK\Big(\frac{T}{\Theta_D}\Big)^3\int_0^{\Theta_D/T}\frac{x^4e^x}{(e^x-1)^2}dx.
$$
The quantity $\Theta_D$ is what we call the {\it Debye temperature}. It  is defined by 
$$
\Theta_D=\frac{\hslash}{K}\sqrt{\lambda_D}.
$$
In particular, it is found that, as $T$ goes to zero, 
\begin{eqnarray*}
C(T)&\sim&\frac{12}{5}\pi^4nK \Big(\frac{T}{\Theta_D}\Big)^3\\
&\sim & \frac{2}{15}\pi^2{\bf V}\Big(\frac{1}{c_l^3}+\frac{2}{c_t^3}\Big)K^4\hslash^{-3}T^3 \quad (\text{in the isotropic case}).
\end{eqnarray*}
This is the $T^3$-{\it law} which agrees well with  the experimental data. 

Debye's model (\ref{debye1}) is still crude since the real shape of the distribution $\varphi(\lambda)$ turns out to be quite different from (\ref{debye1}) in the region of high frequencies.  
As far as the $T^3$-law is concerned, however, we only need the asymptotic property  
\begin{equation}\label{asympto}
\varphi(\lambda)\sim c_0\lambda^{3/2} \quad (\lambda \downarrow 0). 
\end{equation}
To establish this asymptotic behavior in a rigorous way, we shall observe, on the one hand, that $\varphi(\lambda)$ is expressed as the {\it von Neumann trace} of the projection $E(\lambda)$ appearing in the spectral resolution of the operator $D$. On the other hand, we see that $D$ is decomposed into a {\it direct integral}  over the unitary character group of the lattice $L$ (this corresponds to the decomposition of the lattice vibration into independent simple harmonic oscillators). In these discussions, a crucial fact is that $D$ commutes with the action of the lattice on the crystal lattice. 

 A discrete analogue of {\it trace formulae} elucidates a close relation between the function $\varphi$ and  the family of perturbed operators appearing in this direct integral (actually the formula (\ref{internal}) is derived from this observation). Up to this point, we do not require a detailed form of the operator $D$. To proceed further, we must impose  a special condition on the {\it matrix of atomic force constants} which seems appropriate from both the nature of dynamics and the geometric view.  This condition combined with 
 a standard perturbation technique 
allows us to establish (\ref{asympto}) without resorting to Debye's continuum theory. 

Overall, we shall follow, in a slightly different fashion, the journey which physicists usually make to 
perform computation of the specific heat. Besides some mathematical tools and ideas, the main difference is in the use of the terminology in graph theory which has a great advantage:
 not only allows it to avoid the redundancy of suffixes in the formulas, but also naturally brings
 geometric ideas at our disposal. This is the reason why we describe the crystal lattice as a graph.

\section{Crystal lattices}
We shall use the following notations. 
Given an oriented edge $e\in E$ of a graph $X=(V,E)$, we put 
\begin{eqnarray*}
oe&=&\text{the origin of}~e,\\
te&=&\text{the terminus of}~e,\\
\overline{e}&=&\text{the inverse edge of}~e,
\end{eqnarray*}
and write, for $x\in V$,
$$
E_x=\{e\in E;~oe=x\}.
$$

Let $X=(V,E)$ be a crystal lattice realized in ${\bf R}^3$. 
We will always assume that the graph $X$ is connected.
(This assumption is natural if we wish $X$ to imitate a single piece of a real crystal.)
The lattice $L$ acts freely on $X$ 
by graph-automorphisms through the map $\varPhi$. We write the action of $L$ on $V$ and $E$ as $(\sigma,x)\mapsto\sigma x$ and $(\sigma,e)\mapsto \sigma e$, respectively. Hence $\varPhi(\sigma x)=\varPhi(x)+\sigma$.  
We denote by $V_0$ (resp. $E_0$) the quotient set of $V$ (resp. $E$) by the action of $L$. Then a graph structure is induced on $X_0=(V_0,E_0)$ in such a way that the crystal lattice $X$ is an infinite-fold abelian covering graph over $X_0$ with the covering transformation group $L$.
The set $V_0$ is obviously finite. We assume that $E_0$ is also finite (thus we are treating the case of finite range interaction). For the later purpose, we take a fundamental set $\mathcal{F}$ in $V$ for the $L$-action.

Assume that two atoms $x$ and $y$ are of the same kind if they are in the same $L$-orbit, that is, if there exists $\sigma\in L$ such that $y=\sigma x$.  Denote by $m(x)$ the mass of the atom $x$. As a function on $V$, $m$ is $L$-invariant, and hence is regarded as a function on $V_0$. 

 We fix a {\it unit cell}, that is, a fundamental domain for the action of $L$ on ${\bf R}^3$ by translations (for instance, take a fundamental parallelotope). The number of atoms in a unit cell coincides with the number of vertices in $X_0$, which we denote by $n$. We write $m(V_0)=\sum_{x\in V_0}m(x)$, 
 which is the total mass of atoms in the unit cell.

We denote by ${\bf a}\cdot {\bf b}$ the standard inner product on ${\bf R}^3$. Let $L^{*}$ be the dual lattice of $L$ defined by 
$$
L^{*}=\{\eta\in {\bf R}^3;~\eta\cdot \sigma\in {\bf Z}~~ (\sigma\in L)\}.
$$
The lattice $2\pi L^{*}$ is what physicists usually call the {\it reciprocal lattice}. Thus $2\pi\eta$ ~$(\eta\in {\bf R}^3)$ play the role of {\it wavenumber vectors} in the physical context.

\medskip

\noindent{\bf Examples} (1)~ The {\it cubic lattice} $X=(V,E)$ is a crystal lattice with $V={\bf Z}^3\subset {\bf R}^3$. Two vertices $(m_1,m_2,m_3),$ $(n_1,n_2,n_3)$ are joined by an edge if and only if $\sum_{i=1}^3|m_i-n_i|=1$. For the standard lattice $L={\bf Z}^3$ acting on $X$ in a natural manner, the quotient graph $X_0$ is the 3-bouquet graph consisting of a unique vertex with three loop edges.

\smallskip

(2)~ The {\it diamond lattice} is defined as follows. Let $e_1,e_2,e_3$ be the standard basis of ${\bf R}^3$, and let $L$ be the lattice generated by
$
e_1+e_2,~~e_2+e_3,~~e_3+e_1.
$ 
Then put 
$$
V=L\cup \big(L+(1/2,1/2,1/2)\big). 
$$
Two vertices joined by an edge should have the forms 
$$(m_2+m_3,m_3+m_1,m_1+m_2),\quad (n_2+n_3+\frac{1}{2},n_3+n_1+\frac{1}{2}, n_1+n_2+\frac{1}{2})
$$
satisfying one of the following conditions
\begin{eqnarray*}
{\rm (i)}&&m_1=n_1,\quad \quad\quad m_2=n_2,\quad\quad\quad m_3=n_3,\\
{\rm (ii)}&&m_1=n_1+1, \quad  m_2=n_2,\quad\quad\quad m_3=n_3,\\
{\rm (iii)}&&m_1=n_1,\quad\quad\quad m_2=n_2+1,\quad m_3=n_3,\\
{\rm (iv)}&&m_1=n_1,\quad\quad\quad m_2=n_2,\quad\quad\quad m_3=n_3+1.
\end{eqnarray*}
It is easily checked that the quotient graph $X_0$ by the action of $L$ on $X$ is the graph with two vertices joined by 4 multiple edges. 


\section{The equation of motion}

The purpose of this section is to seek the form of the operator $D$. To this end, we impose several conditions on the dynamics of lattice vibrations. The discussion here is rather formal in the sense that we do not specify the domains of linear operators which we introduce. 

The lattice vibrations are supposed to be governed by a potential energy 
$u_{\varPhi}({\bf f})=u(\varPhi+{\bf f})$. Expand it for small $f$ as 
$$
u_{\varPhi}({\bf f})=u_{\varPhi}({\bf 0})+G_{\varPhi}\cdot{\bf f}+\frac{1}{2}K_{\varPhi}{\bf f}\cdot {\bf f}+\cdots,
$$ 
where ${\bf f}\cdot {\bf g}=\sum_{x\in V}{\bf f}(x)\cdot {\bf g}(x)$, and $K_{\varPhi}$ is a linear operator acting on displacements and satisfying 
\begin{equation}
K_{\varPhi}{\bf f}\cdot {\bf g}={\bf f}\cdot K_{\varPhi}{\bf g}.
\end{equation}
The first order term $G_{\varPhi}\cdot{\bf f}$ must be zero, and $K_{\varPhi}{\bf f}\cdot{\bf f}\geq 0$ because of the stability of the equilibrium positions. By omitting higher order terms as usual (i.e. neglecting a coupling produced by ``anharmonic" terms), we write 
$$
u_{\varPhi}({\bf f})=u_{\varPhi}({\bf 0})+\frac{1}{2}K_{\varPhi}{\bf f}\cdot{\bf f},
$$ 
and get the equation of motion 
\begin{equation}\label{motion}
m(x)\frac{d^2{\bf f}}{dt^2}(x)=-{\rm grad}~u_{\varPhi}=-(K_{\varPhi}{\bf f})(x).
\end{equation}
Since the atomic force acts between two atoms $x$ and $y$ if and only if they are joined by an edge, we may write
$$
-(K_{\varPhi}{\bf f})(x)=\sum_{e\in E_x}A_{\varPhi}(e){\bf f}(te)+B_{\varPhi}(x){\bf f}(x),
$$
where $A_{\varPhi}(e)$ and $B_{\varPhi}(x)$ are linear transformations of ${\bf R}^3$.

We impose a somewhat strong assumption on $K_{\varPhi}$. We assume that, if $\varPhi'$ is another realization obtained by a rigid motion of $\varPhi$, then the equation 
$$
m(x)\displaystyle\frac{d^2{\bf f}'}{dt^2}(x)=-(K_{\varPhi'}{\bf f}')(x)
$$
is equivalent to (\ref{motion}) provided that $\varPhi+{\bf f}=\varPhi'+{\bf f}'$.  This property leads to 
$$
K_{\varPhi'}=K_{\varPhi},\quad K_{\varPhi'}(\varPhi-\varPhi')=0.
$$
We write $K=K_{\varPhi},$ $A=A_{\varPhi},$ $B=B_{\varPhi}$. 

Applying $K_{\varPhi'}(\varPhi-\varPhi')=0$ to the case $\varPhi'=\varPhi+v$, ~$(v\in {\bf R}^3)$, 
we get 
$$
\sum_{e\in E_x}A(e)+B(x)=0,
$$
or equivalently
$$
-(K{\bf f})(x)=\sum_{e\in E_x}A(e)\big({\bf f}(te)-{\bf f}(oe)\big),
$$
so that if we put 
\begin{eqnarray*}
D{\bf f}(x)=\frac{1}{m(x)}\sum_{e\in E_x}A(e)\big({\bf f}(te)-{\bf f}(oe)\big),
\end{eqnarray*}
then the equation of motion is given by 
$$
\frac{d^2{\bf f}}{dt^2}=D{\bf f}. 
$$

Applying also $K_{\varPhi'}(\varPhi-\varPhi')=0$ to the case $\varPhi'=U\varPhi$, ~$U\in SO(3)$, we have 
$$
\sum_{e\in E_x}A(e)Uv(e)=\sum_{e\in E_x}A(e)v(e),
$$
where $v(e)=\varPhi(te)-\varPhi(oe)$ which may be regarded as a function on $E_0$ due to the periodicity. Since $\{U-I;~U\in SO(3)\}$ spans $M_3({\bf R})$, 
the space of all real $3\times 3$ matrices (see Lemma \ref{techlemma}  for a proof), we conclude 
$$
\sum_{e\in E_x}A(e)Tv(e)=0
$$
for every $T\in M_3({\bf R})$, or equivalently 
\begin{equation}\label{inv}
\sum_{e\in E_x}A(e)\otimes v(e)=0 \quad (\text{as a tensor}), 
\end{equation}
that is 
$$
\sum_{e\in E_x}A(e)_{ij} v(e)_k=0 
$$
for every $i,j,k$. 

We call $A(e)$ the {\it matrix of atomic force constants}. From the nature of crystals, it is natural to assume that $A(\sigma e)=A(e)$ ~$(\sigma\in L)$ (hence $A(e)$ is regarded as a matrix-valued function on $E_0$). The symmetry condition $K_{\varPhi}{\bf f}\cdot {\bf g}={\bf f}\cdot K_{\varPhi}{\bf g}$ is equivalent to $A(\overline{e})={}^tA(e)$. If $A(e)$ is symmetric, i.e. $A(e)=A(\overline{e})$, then the condition $K{\bf f}\cdot {\bf f}\geq 0$ is equivalent to 
$$
\sum_{e\in E}A(e)\big({\bf f}(te)-{\bf f}(oe)\big)\cdot \big({\bf f}(te)-{\bf f}(oe)\big)\geq 0.
$$

From now on, we assume, together with (\ref{inv}), that $A(e)$ is symmetric and positive definite
(this condition is natural if we assume that the interaction among atoms is a superposition of two-body interactions).

\medskip

\noindent{\bf Example} (1) ~(Monoatomic crystal lattices)\quad This is the case that $X_0$ is a bouquet graph (i.e. the case that the unit cell contains exactly one atom). If $A(e)$ is symmetric, then the condition (\ref{inv}) is satisfied since $v(\overline{e})=-v(e)$. 

(2) ~(The scalar model)\quad  This is the case that $A(e)=a(e)I$ with a positive-valued function $a(e)$ on $E$ such that $a(e)=a(\overline{e})$. If 
\begin{equation}\label{har}
\sum_{e\in E_x}a(e)\big(\varPhi(te)-\varPhi(oe)\big)=0,
\end{equation}
then the condition (\ref{inv}) is satisfied. It should be interesting to point out that $\varPhi$ satisfying (\ref{har}) is a discrete analogue of (vector-valued) harmonic functions, and that $\varPhi$ induces a ``harmonic map" of $X_0$ into the flat torus ${\bf R}^3/L$ (see \cite{kot3}). 

\medskip

We conclude this section with a proof for the following lemma which we have employed to deduce the condition (\ref{inv}). 


\begin{lemma}\label{techlemma} 
Let $T\in M_n({\bf R})$ be a real $n\times n$ matrix. If $n\geq 3$, then there exist $U_1,\ldots,U_N\in SO(n)$ and real scalars $c_1,\ldots,c_N$ such that 
$$
T=c_1(U_1-I)+\cdots +c_N(U_N-I).
$$
\end{lemma}

{\it Proof.}\quad It is enough to prove that, if ${\rm tr}~T(U-I)=0$ for every $U\in SO(n)$, then $T=O$ (the zero matrix). Take a skew-symmetric matrix $S$. Then $e^{tS}\in SO(n)$ for $t\in {\bf R}$, so that differentiating both sides of ${\rm tr}~Te^{tS}={\rm tr}~T$, we find ${\rm tr}~TS=0$. Using 
$$
{\rm tr}~T^{*}S={\rm tr}~ST^{*}=-{\rm tr}~S^{*}T^{*}=-{\rm tr}~TS=0,  
$$
we obtain ${\rm tr}~(T-T^{*})S=0$. Thus $T=T^{*}$. Without loss of generality, we may assume 
$T={\rm diag}(\lambda_1,\ldots,\lambda_n)$, a diagonal matrix. 
For $U=(u_{ij})\in SO(n)$, 
$$
{\rm tr}~T={\rm tr}~TU=\lambda_1u_{11}+\cdots +\lambda_nu_{nn},
$$
so that
$$
\lambda_1+\cdots+\lambda_n=\lambda_1u_{11}+\cdots +\lambda_nu_{nn}.
$$
Applying this to 
\begin{eqnarray*}
U=\begin{pmatrix}
\cos\theta & -\sin\theta & O \\
\sin\theta & \cos\theta & \\
O &  & I_{n-2}
\end{pmatrix},
\end{eqnarray*}
we obtain
$$
\lambda_1+\cdots+\lambda_n=(\lambda_1+\lambda_2)\cos\theta+\lambda_3+\cdots+\lambda_n,
$$
which implies that $\lambda_1+\lambda_2=0$. In the same way, we have $\lambda_i+\lambda_j=0$ 
for $i\neq j$. Therefore we conclude that $\lambda_1=\cdots=\lambda_n=0$, and hence $T=O$ as desired.   $\square$

\medskip
\noindent{\bf Remark}\quad If $n=2$,   then the claim of Lemma \ref{techlemma} is not true.

\section{Continuum limit of a crystal lattice}

The condition (\ref{inv}) was deduced by a rather forcible and formal argument. However, 
this condition turns out to be natural as we shall see below, once we agree that a crystal lattice is a discretization of a (uniform) {\it elastic body}, or the other way around, the {\it continuum limit} of a crystal lattice is an elastic body. We will also see that (\ref{inv}) plays a significant role in the later discussion.

We shall infer what the continuum limit of the crystal lattice $X$ should be by comparing lattice vibrations with {\it elastic waves}. The outcome is used to give a physical meaning to the constant $c_0$, but is not required  to establish the asymptotic of the density of states.

We first recall that, in general, an elastic body is characterized by the {\it mass density} $\rho$ and the 
{\it elastic constant tensor} $C_{\alpha i\beta j}$ satisfying 
$C_{\alpha i\beta j}=C_{i\alpha \beta j}=C_{\alpha i j\beta}=C_{\beta j\alpha i}$ (see \cite{som}). 
An elastic wave ${\bf f}={\bf f}(t, {\bf x})$,
$t\in {\bf R}$, ${\bf x}\in {\bf R}^3$, propagating in the elastic body,  is a solution of the wave equation 
\begin{equation}\label{wave}
\rho\frac{\partial^2 {\bf f}}{\partial t^2}=\sum_{i,j=1}^3\frac{\partial}{\partial x_i}\Big(A_{ij}\frac{\partial {\bf f}}{\partial x_j}\Big),
\end{equation}
where, for each $i, j$, $A_{ij}$ is a matrix-valued function on ${\bf R}^3$ defined by 
$$
(A_{ij})_{\alpha\beta}=C_{\alpha i\beta j}\quad (\text{the $(\alpha,\beta)$-component of the matrix $A_{ij}$}).
$$
We easily see that ${}^tA_{ij}=A_{ji}$.

We assume that the elastic body obtained as the continuum limit is {\it uniform} in the sense that $\rho$ and $A_{ij}$ are constant. Symmetrizing $A_{ij}$ if necessary, we may assume $A_{ij}=A_{ji}$. 

It is natural, from the nature of mass density, to put $\rho=m(V_0)/{\bf V}$. To surmise the form of the matrix $A_{ij}$, take  a smooth function ${\bf f}:{\bf R}\times {\bf R}^3\rightarrow {\bf R}^3$, and define ${\bf f}_{\delta}:{\bf R}\times V\rightarrow {\bf R}^3$ by setting 
$$
{\bf f}_{\delta}(t,x)={\bf f}\big(\delta t,\delta \varPhi(x)\big)
$$
(thus ${\bf f}_{\delta}$ is a discretization of ${\bf f}$ with respect to the space variable). Then 
\begin{eqnarray*}
-(K{\bf f}_{\delta})(t,x)&=&\sum_{e\in E_x}A(e)\big[{\bf f}(\delta t,\delta \varPhi(te))-{\bf f}(\delta t,\delta \varPhi(oe))\big]\\
&=&\delta \sum_{i=1}^3\sum_{e\in E_x}v(e)_i A(e)\frac{\partial {\bf f}}{\partial x_i}\big(\delta t,\delta\varPhi(x)\big)\\
~&+&\frac{1}{2}\delta^2\sum_{i,j=1}^3\sum_{e\in E_x}v(e)_iv(e)_jA(e)\frac{\partial^2{\bf f}}{\partial x_i\partial x_j}\big(\delta t,\delta\varPhi(x)\big)+\cdots,
\end{eqnarray*}
where $v(e)=\big(v(e)_1,v(e)_2,v(e)_3\big)$. Under the condition (\ref{inv}), the first term vanishes. Therefore, we have, for a sequence $\{x_{\delta}\}$ in $V$ with  
$\lim_{\delta\downarrow 0}\delta\varPhi(x_{\delta})={\bf x}$, and a sequence $\{t_{\delta}\}$ in ${\bf R}$ with $\lim_{\delta\downarrow 0}\delta t_{\delta}=t$,
\begin{eqnarray*}
\lim_{\delta\downarrow 0}-\delta^{-2}\big(K{\bf f}_{\delta}\big)(t_{\delta},x_{\delta})
 =\frac{1}{2}
\sum_{i,j=1}^3\sum_{e\in E_{0,x}}v(e)_iv(e)_j A(e)\frac{\partial^2{\bf f}}{\partial x_i\partial x_j}(t,{\bf x}),
\end{eqnarray*}
 provided that $\{x_{\delta}\}$ is in the orbit containing $x\in \mathcal{F}$. On the other hand, 
$$
\lim_{\delta\downarrow 0}\delta^{-2}m(x_{\delta})\frac{d^2 {\bf f}_{\delta}}{dt^2}(t_{\delta},x_{\delta})=m(x)\frac{\partial^2 {\bf f}}{\partial t^2}(t,{\bf x}).
$$
Thus, taking the sum over the fundamental set $\mathcal{F}$, and dividing by ${\bf V}$, we may presume that the equation of motion approaches the equation
$$
\rho\frac{\partial^2 {\bf f}}{\partial t^2}=\frac{1}{2{\bf V}}
\sum_{i,j=1}^3\sum_{e\in E_{0}}v(e)_iv(e)_j A(e)\frac{\partial^2 {\bf f}}{\partial x_i\partial x_j}
$$
as the mesh of the lattice becomes finer. This implies that the (symmetrized) elastic constant tensor of the elastic body corresponding to our crystal lattice is given by  
$$
A_{ij}=\frac{1}{2{\bf V}}\sum_{e\in E_{0}}v(e)_iv(e)_j A(e). 
$$

\medskip 

\noindent{\bf Remark}\quad The differential operator (the {\it elastic Laplacian})
$$
\rho^{-1}\sum_{i,j=1}^3\frac{\partial}{\partial x_i}\Big(A_{ij}\frac{\partial}{\partial x_j}\Big)
$$
can be written as $-d^{*}d$, where 
$$
d: A^0({\bf R}^3,{\bf R}^3) \longrightarrow A^1({\bf R}^3,{\bf R}^3)
$$
is the exterior differentiation 
acting on ${\bf R}^3$-valued differential forms, that is, 
$$
d{\bf f}=\sum_{i=1}^3 \frac{\partial {\bf f}}{\partial x_i}dx_i.
$$
The operator $d^{*}$ is the formal adjoint of $d$ with respect to the inner products on $A^i({\bf R}^3,{\bf R}^3)$ ~$(i=0,1)$ defined by 
\begin{eqnarray*}
&&\langle {\bf f},{\bf g}\rangle =\int_{{\bf R}^3} {\bf f}\cdot {\bf g} ~\rho~d{\bf x}\quad \quad ({\bf f},{\bf g}\in A^0({\bf R}^3,{\bf R}^3)),\\
&&\langle \omega,\eta\rangle=\int_{{\bf R}^3}\sum_{i,j=1}^3 A_{ij}\omega_j\cdot \eta_i~ d{\bf x} \quad (\omega,\eta\in A^1({\bf R}^3,{\bf R}^3)),
\end{eqnarray*}
where $\omega=(\omega_1,\omega_2,\omega_3),$ $\eta=(\eta_1,\eta_2,\eta_3)$.

\medskip

We now return to the difference operator $D$. We easily see that 
$D$ is an $L$-equivariant linear operator of $C(V,{\bf C}^3)$, the space of ${\bf C}^3$-valued functions on $V$. If we define the Hilbert space $\ell^2(V,m)$ by \begin{eqnarray*}
\ell^2(V,m)=\{{\bf f}\in C(V,{\bf C}^3);~\|{\bf f}\|^2: =\sum_{x\in V}{\bf f}(x)\cdot \overline{{\bf f}(x)}m(x)<\infty\},
\end{eqnarray*}
then $D$ restricted to the subspace $\ell^2(V,m)$ is a  bounded self-adjoint operator of $\ell^2(V,m)$. We call $D$ the {\it discrete elastic Laplacian}. This naming is justified by the expression
$$
D=-d^{*}d,
$$
where $d:\ell^2(V,m)\rightarrow \ell^2(E,A)$, a discrete analogue of the exterior differentiation (or the 
{\it coboundary operator} in cohomology theory), is defined by 
$$
d{\bf f}(e)={\bf f}(te)-{\bf f}(oe),
$$
and 
\begin{eqnarray*}
\ell^2(E,A)&=&\{\eta:E\rightarrow {\bf C}^3;~\eta(\overline{e})=-\eta(e),\\
&&\quad \quad ~ \|\eta\|^2:=\frac{1}{2}\sum_{e\in E}A(e)\eta(e)\cdot \overline{\eta(e)}<\infty\}.
\end{eqnarray*}
Indeed, the explicit expression for  $d^{*}$ is given by
\begin{equation}\label{adjoint}
(d^{*}\omega)(x)=-\frac{1}{m(x)}\sum_{e\in E_x}A(e)\omega(e).
\end{equation}
This is checked by the following computation.
\begin{eqnarray*}
\langle d{\bf f},\omega\rangle&=&\frac{1}{2}\sum_{e\in E}A(e)\big({\bf f}(te)-{\bf f}(oe)\big)\cdot \overline{\omega(e)}\\
&=&\frac{1}{2}\sum_{e\in E}\big({\bf f}(te)-{\bf f}(oe)\big)\cdot \overline{{}^tA(e)\omega(e)}\\
&=& \frac{1}{2}\sum_{e\in E}{\bf f}(te)\cdot \overline{A(\overline{e})\omega(e)}-
\frac{1}{2}\sum_{e\in E}{\bf f}(oe)\cdot \overline{A(e)\omega(e)}\\
&=&-\sum_{e\in E}{\bf f}(oe)\cdot \overline{A(e)\omega(e)},
\end{eqnarray*}
where we have used the assumption ${}^tA(e)=A(e)$. The last term is written as 
\begin{eqnarray*}
-\sum_{x\in V}{\bf f}(x)\cdot \Big(\frac{1}{m(x)}\sum_{e\in E_x}\overline{A(e)\omega(e)}\Big)m(x),
\end{eqnarray*} 
from which (\ref{adjoint}) follows.

Therefore the operator $D$ is not only a discretization of the elastic Laplacian, 
but also  its conceptual analogue.  

\section{Hamiltonian formalism for lattice vibrations}
This section is devoted to a brief explanation for the quantization of lattice vibrations which is performed in a slightly different way from the current one in the physical literature. To this end, we shall start with the Hamiltonian formalism for lattice vibrations. 

We put $S=\ell^2(V,m)$, and denote by $\langle \cdot,\cdot\rangle$ the inner product on the Hilbert space $S$. We shall regard $S$ as a {\it symplectic vector space} with the symplectic form $\omega$ defined by 
$$
\omega(u,v)={\rm Im}\langle u,v \rangle \quad  (u,v\in S)
$$
(${\rm Im}~z$ denotes the imaginary part of $z\in {\bf C}$). 
Define the Hamiltonian $H$ by 
$$
H(u)=\frac{1}{2}\langle\sqrt{-D}u,u\rangle.
$$
Then the Hamiltonian equation for $H$ is given by 
$$
\frac{du}{dt}=-\sqrt{-1}\sqrt{-D}u,
$$
which is obviously equivalent to the equation $\displaystyle\frac{d^2{\bf f}}{dt^2}=D{\bf f}$. 
(The equivalence can be established, for example, by the relation ${\bf f}={\rm Re}\, u$).

We wish to quantize the Hamiltonian system $(S,\omega,H)$. To avoid the difficulty  arising from the infinite-dimensionality of $S$, we shall decompose $(S,\omega,H)$ into a direct integral of finite dimensional Hamiltonian systems. The idea, which essentially dates back to Bloch's work on periodic Schr\"{o}dinger operators and has been taken up in a different manner by physicists, is to use the irreducible decomposition of the regular representation $\rho_r$ of $L$ on the Hilbert space 

\noindent
$\ell^2(L)=\{f:L\rightarrow {\bf C};~\sum_{\sigma\in L}|f(\sigma)|^2<\infty\}$:
\begin{equation}\label{dir}
(\rho_r,\ell^2(L))=\int_{\widehat{L}}^{\oplus}(\chi,{\bf C})d\chi,
\end{equation}
where 
$\widehat{L}$ is the unitary character group of $L$, and $d\chi$ denotes 
the  Haar measure on $\widehat{L}$, which is normalized so that
$$
\int_{\widehat{L}} d\chi = 1.
$$

Similarly to (\ref{dir}), we may construct a direct integral decomposition
\begin{equation}\label{dir1}
(S,\omega,H)=\int_{\widehat{L}}^{\oplus}(S_{\chi},\omega_{\chi},H_{\chi})d\chi.
\end{equation}
Namely, we take 
\begin{eqnarray*}
&&S_{\chi}=\{u:V\rightarrow {\bf C}^3;~u(\sigma x)=\chi(\sigma)u(x)\},\\
&&\omega_{\chi}(u,v)={\rm Im}\langle u,v\rangle_{\chi},\\
&& H_{\chi}(u)=\frac{1}{2}\langle \sqrt{-D_{\chi}}u,u\rangle_{\chi},
\end{eqnarray*}
where 
$$
\langle u,v\rangle_{\chi}=\sum_{x\in \mathcal{F}}u(x)\cdot \overline{v(x)}m(x)
$$
is the scalar product in $S_\chi$.
Note that ${\rm dim}~S_{\chi}=3n$,  and $S_\chi$ is an invariant subspace
(in fact, even an eigenspace) of the action of $L$ in $C(V,{\bf C}^3)$
(the space of all ${\bf C}^3$-valued fuctions on $V$). Late we will sometimes refer 
to the elements of $S_\chi$ as {\it Bloch functions}, and to the components
of different objects in the direct integral decomposition above as the 
{\it Bloch components}.

Note that the action of $L$ in $S_{\chi}$ is not irreducible
(unlike the corresponding subspace for the action of $L$  in $\ell^2(L)$). 
The isometry between $S$ and the direct integral
$$
\int_{\widehat{L}}^{\oplus}S_{\chi}~d\chi
$$
is given as follows. For $u\in S$ with finite support, define $u_{\chi}\in S_{\chi}$ by 
\begin{equation}\label{E:u-chi}
u_{\chi}(x)=\sum_{\sigma\in L}\chi(\sigma)^{-1}u(\sigma x).
\end{equation}
Then the extension of the correspondence $u\mapsto \{u_{\chi}\}$ gives rise to the desired isometry. 
The inverse isometry is
\begin{equation}\label{E:s-chi}
\left\{s_\chi\in S_\chi; \;\chi\in {\widehat{L}}\right\}  \mapsto u(x)=\int_{\widehat{L}} s_\chi(x) d\chi,
\end{equation}
where we assume that the function $\chi\mapsto s_\chi$ is square-integrable on $\widehat{L}$.

The ``twisted" operator $D_{\chi}$ is defined to be the restriction of 
$D:C(V,{\bf C}^3)\longrightarrow C(V,{\bf C}^3)$ to $S_{\chi}\subset C(V,{\bf C}^3)$.  
It should be pointed out that, in physical terms, the direct integral decomposition (\ref{dir1}) 
corresponds to the ``sum" over {\it wavenumber vectors}.

Since $d$ and $d^{*}$ are also decomposed as 
\begin{eqnarray*}
&&d=\int_{\widehat{L}}^{\oplus}d_{\chi}~d\chi,\\
&&d^{*}=\int_{\widehat{L}}^{\oplus}(d^{*})_{\chi}~d\chi,
\end{eqnarray*}
and $(d^{*})_{\chi}=(d_{\chi})^{*}$, we have $D_{\chi}=-d_{\chi}^{*}d_{\chi}$, and hence $D_{\chi}\leq 0$. Furthermore, if $\chi\neq {\bf 1}$~ (${\bf 1}$ being the trivial character), then $D_{\chi}<0 $. 
 Note that $0$ is an eigenvalue of $-D_{\bf 1}$ of multiplicity three whose eigenfunctions are constant,  
 because we assumed $X$ to be connected (hence $X_0$ is connected too).
We enumerate the eigenvalues of $-D_{\chi}$ as 
$$
0 \leq \lambda_1(\chi)\leq \lambda_2(\chi)\leq \cdots\leq \lambda_{3n}(\chi).
$$
 The functions $\lambda_i(\chi)$ are continuous on $\widehat{L}$, and the first three eigenvalues $\lambda_1(\chi),$ $\lambda_2(\chi),$ $\lambda_3(\chi)$ are perturbations of the eigenvalue $0=\lambda_1({\bf 1})=\lambda_2({\bf 1})=\lambda_3({\bf 1})$ of $-D_{\bf 1}$, which 
are said to be {\it acoustic branches}, while other eigenvalues are said to be the {\it optical branches}.  

\medskip

\noindent{\bf Remark.}\quad The spectrum $\sigma(-D)$ coincides with 
$$
\bigcup_{k=1}^{3n}~ {\rm Image}~\lambda_k.
$$
Therefore $-D$ has a {\it band spectrum} in the sense that $\sigma(-D)$ is a union of finitely many closed intervals. In the case of monoatomic lattices, we only have acoustic branches, and $\sigma(-D)$ is an interval. 

\medskip

Choose an orthonormal basis $e_1,\ldots,e_{3n}$ of $S_{\chi}$ such that 
$$
-D_{\chi}e_i=\lambda_i(\chi)e_i.
$$
Then the ${\bf R}$-basis $\{e_1,\ldots,e_{3n},\sqrt{-1}e_1,\ldots,\sqrt{-1}e_{3n}\}$ yields a canonical linear coordinate system $(p_1,\ldots,p_{3n},$ $q_1,\ldots,q_{3n})$ by setting 
$$
u=\sum_{i=1}^{3n}(q_i+\sqrt{-1}p_i)e_i.
$$ 
In terms of this coordinate system, the Hamiltonian $H_{\chi}$ is expressed as 
$$
H_{\chi}=\frac{1}{2}\sum_{i=1}^{3n}\sqrt{\lambda_i(\chi)}\big(p_i^2+q_i^2\big).
$$
Note that $H_{\chi,i}=\displaystyle\frac{1}{2}\sqrt{\lambda_i(\chi)}\big(p_i^2+q_i^2\big)$ is the Hamiltonian of the simple harmonic oscillator with the frequency $\sqrt{\lambda_i(\chi)}/2\pi$. Therefore the quantized Hamiltonian for $H_{\chi}$ is given by 
\begin{eqnarray*}
\widehat{H}_{\chi}=\frac{1}{2}\sum_{i=1}^{3n}\sqrt{\lambda_i(\chi)}\Big(-\hslash^2\frac{\partial^2}{\partial q_i^2}+q_i^2\Big)=\sum_{i=1}^{3n}\widehat{H}_{\chi,i}.
\end{eqnarray*}
The standard fact tells that the spectrum of the quantized Hamiltonian $\widehat{H}_{\chi,i}$  ~$(\chi\neq {\bf 1})$ consists of simple eigenvalues
$$
E_{\chi,i,k}=\hslash \sqrt{\lambda_i(\chi)}\Big(k+\frac{1}{2}\Big) \quad (k=0,1,2,\ldots).
$$
The ``quantum particle" with the energy $E_{\chi,i,k}$ is said to be a {\it phonon}.

\section{Computation of the internal energy}

Following a general recipe in quantum statistical mechanics (more precisely,
 the Bose-Einstein statistics for phonons), we define the partition function 
 for $\widehat{H}_{\chi,i}$  by 
\begin{eqnarray*}
Z_{\chi,i}(T)=\sum_{k=0}^{\infty}e^{-E_{\chi,i,k}/KT}=
\frac{e^{-\hslash \sqrt{\lambda_i(\chi)}/2KT}}{1-e^{-\hslash \sqrt{\lambda_i(\chi)}/KT}}.
\end{eqnarray*}
Then the internal energy of $\widehat{H}_{\chi,i}$ for the Gibbs distribution
$$
p_k=Z_{\chi,i}(T)^{-1}e^{-E_{\chi,i,k}/KT} \quad (k=0,1,2,\ldots)
$$
is given by 
$$
U_{\chi,i}(T)=\sum_{k=0}^{\infty} E_{\chi,i,k}p_k=
\hslash \sqrt{\lambda_i(\chi)}\left[\frac{1}{2}+\frac{1}{e^{\hslash \sqrt{\lambda_i(\chi)}/KT}-1}
\right]. 
$$

In view of the additive property of the internal energy, it is natural to define the internal energy of the lattice vibration (per  unit cell) by 
\begin{eqnarray*}
U(T)=\int_{\widehat{L}}\sum_{i=1}^{3n}U_{\chi,i}(T)~d\chi,
\end{eqnarray*}
which is expressed as
\begin{eqnarray*}
\frac{\hslash}{2}\int_{\widehat{L}}{\rm tr}~\sqrt{-D_{\chi}}~d\chi+
\int_{\widehat{L}}{\rm tr}~\frac{\hslash \sqrt{-D_{\chi}}}{e^{\hslash\sqrt{-D_{\chi}}/KT}-1}~d\chi
= U_0+U_1(T).
\end{eqnarray*}

To transform this formula into one in terms of the density of states, we employ the notion 
of {\it von Neumann trace} (or $L$-trace; see \cite{atiyah}), which is defined, 
for a $L$-equivariant bounded linear operator $T:S\rightarrow S$, as 
\begin{equation*}
{\rm tr}_{L}~T=\sum_{x\in \mathcal{F}}{\rm tr}~t(x,x)m(x),
\end{equation*}
where $t(x,y)$ is the kernel function (matrix) of $T$, that is,
\begin{equation}\label{E:matrix}
T{\bf f}(x)=\sum_{y\in V}t(x,y){\bf f}(y)m(y).
\end{equation}
Note that  $t(x,y)$ is a $3\times 3$ complex matrix for fixed $x$ and $y$, and the matrix-valued 
function $x,y\mapsto t(x,y)$ is
invariant under the diagonal action of $L$ on $V\times V$ given by $\sigma (x,y)=(\sigma x, \sigma y)$.  
In particular, the scalar function $x\mapsto {\rm tr}~t(x,x)m(x)$ is $L$-invariant on $V$, hence the trace
${\rm tr}_L~T$ in the definition above does not depend upon the choice of the fundamental domain
$\mathcal{F}$.

We should point out that the notion of $L$-trace is introduced in a more general setting, say in {\it von Neumann algebras} of type II$_1$.  

The $L$-trace is related to the direct integral decomposition. The following Lemma
is known  in the theory of operator algebras (in a more general context),
but we will supply an elementary proof for convenience of the reader. 

\begin{lemma}\label{L:trace} {\rm (Trace Decomposition Formula)}
Let $T:S\to S$ be a bounded linear operator commuting with the action of $L$, and 
$$
T=\int_{\widehat{L}}^{\oplus}T_{\chi}~d\chi
$$
is its direct integral decomposition. Then
\begin{equation}\label{E:trace}
{\rm tr}_{L}~T=\int_{\widehat{L}}{\rm tr}~T_{\chi}~d\chi.
\end{equation}
\end{lemma}

\begin{proof}
First we will introduce some notations.
Let us identify the unitary character group $\widehat{L}$ with the torus 
$$
J_{L}={\bf R}^3/L^{*}
$$
via the correspondence
$$
\chi\in {\bf R}^3 \mapsto \chi'
\in \widehat{L},
$$
where 
$$
\chi'(\sigma)=\exp\big(2\pi \sqrt{-1}\chi\cdot\sigma \big),
$$
so that a small neighborhood $U({\bf 1})$ of ${\bf 1}$ in $\widehat{L}$ 
is identified with a neighborhood $U(0)$ of $0$ in ${\bf R}^3$. 

We will present $T$ in terms of operators
$$
T_\chi^0: C(V_0,{\bf C}^3)\to C(V_0,{\bf C}^3),
$$ 
where $C(V_0, {\bf C}^3)$
is the Hilbert space of all ${\bf C}^3$-valued functions on $V_0$,
with the scalar product
$$
\langle u,v\rangle = \sum_{x\in V_0} u(x)\cdot  \bar v(x) m(x),
$$
 $\dim_{\bf C} C(V_0,{\bf C}^3)=3n$. 
The operator $T_\chi^0$ 
is defined
as $T_\chi^0=U_\chi^{-1} T_\chi U_\chi$, where
$U_\chi: C(V_0, {\bf C}^3)  \to  S_\chi$ is a unitary operator, given by
$$
 (U_\chi {\bf f})(x) =s_\chi(x) =e^{2\pi \sqrt{-1}\chi \cdot \varPhi(x)} {\bf f}(\pi(x)), 
 $$
where $x\in V$, $\pi(x)$ is the canonical projection of $x$ to $V_0=V/L$.
It is easy to check that $s_\chi\in S_\chi$. Indeed,
\begin{eqnarray*}
s_\chi(\sigma x)&=&\exp
\big(2\pi \sqrt{-1} \chi\cdot\varPhi(\sigma x)
\big){\bf f}(\pi(\sigma x))\\
&=&\exp
\big(2\pi \sqrt{-1} \chi\cdot(\varPhi(x)+\sigma)
\big)
{\bf f}(\pi(x))\\
&=& \exp\big(2\pi \sqrt{-1}\chi\cdot\sigma \big)s_\chi(x),
\end{eqnarray*}
as desired. The inverse operator
$$
U_\chi^{-1}: S_\chi \to C(V_0, {\bf C}^3)
$$
is given by
$$
(U_\chi^{-1} s_\chi)(\pi(x)) =  e^{-2\pi \sqrt{-1}\chi \cdot \varPhi(x)} s_\chi(x),
$$
where $x\in V$. (The left hand side is well defined because 
the right hand side is easily seen to be $L$-invariant.)

The advantage of using the operators $T_\chi^0$
(compared with $T_\chi$) is that they act in the same space
$C(V_0, {\bf C}^3)$ for all $\chi\in \widehat{L}$.

Now let us present the operator $T_\chi^0$ by its kernel function (matrix)
$t_\chi^0=t_\chi^0(\bar x, \bar y)$, where $\bar x, \bar y\in V_0$, so that
$$
(T_\chi^0 {\bf f})(\bar x)= \sum_{\bar y\in V_0} t_\chi^0(\bar x, \bar y) {\bf f}(\bar y) m(\bar y).
$$
 
 Let us choose a function $u\in S$ with a finite support, and find its Bloch
 components $u_\chi$  ($\chi\in \widehat{L}$), defined by \eqref{E:u-chi},
 which can be also rewritten in the form
 $$
 u_\chi(x) = \sum_{\sigma\in L} e^{-2\pi\sqrt{-1}\chi\cdot \sigma} u(\sigma x).
 $$
 The Bloch components of $Tu$ are  
 $(Tu)_\chi = T_\chi u_\chi = U_\chi T_\chi^0 U_\chi^{-1}u_\chi$.
 Let us calculate them explicitly. We have
 \begin{eqnarray*}
 &(U_\chi^{-1} u_\chi)(x) = e^{-2\pi\sqrt{-1}\chi\cdot\varPhi(x)} u_\chi(x)
 &= \sum_{\sigma\in L} e^{-2\pi\sqrt{-1}\chi\cdot(\varPhi(x)+\sigma)} u(\sigma x)\\
&&=  \sum_{\sigma\in L} e^{-2\pi\sqrt{-1}\chi\cdot \varPhi(\sigma x)} u(\sigma x),
 \end{eqnarray*}
 which is $L$-invariant (hence well defined on $V_0$). Further, we find
 \begin{eqnarray*}
&(T_\chi^0 U_\chi^{-1} u_\chi)(\pi(x))
&= \sum_{y\in \mathcal{F}, \sigma \in L} t_\chi^0(\pi(x), \pi(y))
e^{-2\pi\sqrt{-1}\chi\cdot \varPhi(\sigma y)} u(\sigma y) m(y)\\
&&= \sum_{y\in V} t_\chi^0(\pi(x), \pi(y))
e^{-2\pi\sqrt{-1}\chi\cdot \varPhi(y)} u(y) m(y).
 \end{eqnarray*}
 Therefore,
\begin{eqnarray*}
&(T_\chi u_\chi)(x) &= (U_\chi T_\chi^0 U_\chi^{-1} u_\chi)(\pi(x))\\
&&= \sum_{y\in V} t_\chi^0(\pi(x), \pi(y))
e^{2\pi\sqrt{-1}\chi\cdot (\varPhi(x) - \varPhi(y))} u(y) m(y).
 \end{eqnarray*}
 Taking into account \eqref{E:s-chi}, we see that 
 \begin{eqnarray*}
&(Tu)(x) &= \int_{\widehat{L}}(T_\chi u_\chi)(x) d\chi\\
&&= \int_{\widehat{L}}\sum_{y\in V} t_\chi^0(\pi(x), \pi(y))
e^{2\pi\sqrt{-1}\chi\cdot (\varPhi(x) - \varPhi(y))} u(y) m(y) d\chi.
 \end{eqnarray*}
 This means that the kernel function of $T$ is given by the formula
 \begin{equation}\label{E:kernel}
 t(x,y)=
 \int_{\widehat{L}}  e^{2\pi\sqrt{-1}\chi\cdot (\varPhi(x) - \varPhi(y))}
 t_\chi^0(\pi(x), \pi(y)) d\chi.
 \end{equation}

 Note that by the general properties of direct integrals
 $$
 \|T: S\to S\|= {\rm ess}\,\sup\{\|T_\chi\|;\; \chi\in \widehat{L}\}
 = {\rm ess}\,\sup\{\|T_\chi^0\|;\; \chi\in \widehat{L}\}.
 $$
 In particular, $T$  is bounded if and only if the norms $\|T_\chi\|=\|T_\chi^0\|$
 are a.e. uniformly bounded for $\chi\in \widehat{L}$, or, equivalently,
 the matrix elements $t_{\chi}^0(\bar x, \bar y)$ are uniformly bounded
a.e. in $\chi$ for all $\bar x, \bar y\in V_0$. 

Now taking $y=x$ in \eqref{E:kernel}, applying matrix trace
(to the $3\times 3$ matrices in both sides) and summing 
over $x\in \mathcal{F}$, we obtain
\begin{eqnarray*}
{\rm tr}_L\; T = \sum_{x\in \mathcal{F}} {\rm tr}\; t(x,x)
=  \int_{\widehat{L}} \sum_{x\in \mathcal{F}} {\rm tr}\; t_\chi^0(\pi(x), \pi(x)) d\chi\\
= \int_{\widehat{L}}  {\rm tr}\; T_\chi^0 d\chi
= \int_{\widehat{L}}  {\rm tr}\; T_\chi d\chi,
\end{eqnarray*}  
which ends the proof. 
 \end{proof}

 \begin{remark}
 The kernel representation formula \eqref{E:kernel}
 also allows to provide a relation between the smoothness
 of the function $\chi\mapsto T_{\chi}^0$ and the off-diagonal decay
 the kernel $t(x,y)$, i.e. its decay as $|\varPhi(x) - \varPhi(y)|\to \infty$.
 More precisely, let us say that a bounded linear operator $T:S\to S$
 (commuting with the action of $L$) is {\it smooth},
 if the matrix elements $t_\chi^0(\bar x, \bar y)$
are in $C^\infty(\widehat{L})$ with respect to $\chi$,
for any $\bar x, \bar y\in V_0$.  If $T$ is smooth, then,
integrating by parts,
we can for any positive integer $N$ rewrite \eqref{E:kernel} as follows:
 \begin{eqnarray*}\label{E:kernel-N}
& t(x,y)=
 (1+|\varPhi(x)-\varPhi(y)|^2)^{-N}\times\\
& \int_{\widehat{L}}  e^{2\pi\sqrt{-1}\chi\cdot (\varPhi(x) - \varPhi(y))}
\left[ (1-\frac{1}{2\pi}\Delta_\chi)^Nt_\chi^0(\pi(x), \pi(y))\right] d\chi,
 \end{eqnarray*}
where $\Delta_\chi$ denotes the Laplacian with respect to $\chi$.
It follows, that
\begin{equation}\label{E:kernel-est}
|t(x,y)|\le C_N(1+|\varPhi(x) - \varPhi(y)|^2)^{-N}, \quad N=0,1,2,\dots.
\end{equation}

Vice versa, if these estimates hold, then we can calculate $T_\chi^0$ as 
the operator $U_\chi^{-1} T U_\chi$, identifying elements of $C(V_0, {\bf C}^3)$
with $L$-invariant ${\bf C}^3$-valued functions ${\bf f}$ on $V$, and then
applying $T$ to the Bloch function $U_\chi{\bf f}$ using the kernel (matrix)
representation of $T$ in terms of $t(x,y)$ (see \eqref{E:matrix}). This leads to the
expression
\begin{eqnarray*}
&(U_\chi^{-1} T U_\chi{\bf f})(x)
&= \sum_{y\in V} e^{- 2\pi \sqrt{-1}\chi\cdot (\varPhi(x)-\varPhi(y))} t(x,y)m(y) {\bf f}(y)\\
 &&= \sum_{y\in \mathcal{F}} \left[\sum_{\sigma\in L} 
 e^{- 2\pi \sqrt{-1}\chi\cdot (\varPhi(x)-\varPhi(\sigma y))} t(x,\sigma y)
 \right] m(y) {\bf f}(y),
\end{eqnarray*}
where the sums converge due to \eqref{E:kernel-est}.
The latter expression is equivalent to the following presentation 
(which is the inversion formula to \eqref{E:kernel})
for the matrix $t_\chi^0$ :
\begin{equation}\label{E:inversion}
t_\chi^0(\pi(x),\pi(y)) = \sum_{\sigma \in L} 
e^{-2\pi\sqrt{-1}\chi\cdot(\varPhi(x)-\varPhi(\sigma y))} t(x,\sigma y),
\end{equation}
where the left hand side is well defined because the right hand side is
$L\times L$-invariant on $V\times V$, i.e $L$-invariant separately in $x$ and $y$.
 We can also differentiate \eqref{E:inversion} with respect to $\chi$
termwise ,  which implies 
the smoothness of the function $\chi\mapsto T_\chi^0$. \hskip2.3in $\square$
\end{remark}

\begin{remark}
From \eqref{E:inversion} we also deduce
$$
{\rm tr}\; T_\chi = {\rm tr}\; T_\chi^0
= \sum_{x\in \mathcal{F}}\sum_{\sigma \in L} 
e^{-2\pi\sqrt{-1}\chi\cdot(\varPhi(x)-\varPhi(\sigma x))} {\rm tr}\; t(x,\sigma x),
$$
which is an analogue of the trace formula (see \cite{sun1}).  
\end{remark}

\medskip
 Let
$$
-D=\int \lambda~dE(\lambda)
$$
be the spectral resolution of $-D$. Then the {\it integrated density of states} is defined by 
$$
\varphi(\lambda)={\rm tr}_{L}~E(\lambda).
$$
We find that 
$\varphi$ is a nondecreasing function, and  
\begin{eqnarray*}
\varphi(\lambda)=\begin{cases}
0 &  (\lambda<0)\\
3n & (\lambda > \|D\|)
\end{cases},
\end{eqnarray*}
so that 
$$
\int_0^{\infty}~d\varphi(\lambda)=3n. 
$$

We should note that, when we take a decreasing sequence of lattices 
$$
L=L_0\supset L_1\supset L_2\supset \cdots
$$
with $\bigcap_{i=0}^{\infty} L_i=\{0\}$, the function $\varphi(\lambda)$ coincides with the limit 
\begin{eqnarray*}
\lim_{i\to\infty} \left[\#(L/L_i)\right]^{-1} \varphi_i(\lambda)
\end{eqnarray*}
at continuity points of $\varphi$, where $\varphi_i$ is the counting function of eigenvalues for the periodic boundary value problem:
\begin{eqnarray*}
&&-D{\bf f}=\lambda {\bf f}, \\
&& {\bf f}(\sigma x)={\bf f}(x) \quad (\sigma\in L_i)
\end{eqnarray*}
(see \cite{sun1}, \cite{shu}), that is, if $\lambda_1\leq \cdots \leq \lambda_N$ ~$(N=\#(V/L_i))$ are the eigenvalues for the boundary problem above, then $\varphi_i(\lambda)=\#\{\lambda_i\leq \lambda\}$. Thus the function $\varphi$ deserves to be called the (integrated) density of states.

Now using 
\begin{eqnarray}
&& f(-D)=\int_{\widehat{L}}f(-D_{\chi})~d\chi,\nonumber\\
&&\int_{\widehat{L}}{\rm tr}~f(-D_{\chi})d\chi={\rm tr}_{L}~f(-D)=
\int f(\lambda)d\varphi(\lambda),\label{trace}
\end{eqnarray}
we get 
\begin{eqnarray*}
&&U_0=\frac{\hslash}{2}\int_{0}^{\infty}\sqrt{\lambda}~d\varphi(\lambda),\\
&&U_1(T)=\int_0^{\infty}\frac{\hslash \sqrt{\lambda}}{e^{\hslash \sqrt{\lambda}/KT}-1}d\varphi(\lambda),\\
&&C(T)=\frac{\partial U_1}{\partial T}=\int_0^{\infty}\frac{\frac{\hslash^2 \lambda}{KT^2}e^{\hslash \sqrt{\lambda}/KT}}{\big(e^{\hslash \sqrt{\lambda}/KT}-1\big)^2}d\varphi(\lambda).
\end{eqnarray*}
$U_0$ is what we call the {\it zero-point energy}, which indicates that vibrations in a solid persist even at the absolute zero of temperature, a peculiar feature of quantum mechanics. 

\medskip 

\noindent{\bf Remark.} (1)\quad It is straightforward to see that 
$$
{\rm supp}~d\varphi={\rm spectrum}(-D).
$$ 

(2)\quad 
We may consider integrated densities of states for other boundary conditions. From the {\it abelian} nature of the crystal lattice, however, it follows that the notion of density of states is independent of the particular boundary conditions imposed. 

(3)\quad The partition function for the Hamiltonian $H_{\chi,i}$ in the classical statistical mechanics is given by 
\begin{eqnarray*}
Z_{\chi,i}(T)&=&\int_{{\bf R}^2}\exp\Big(-\frac{H_{\chi,i}(p_i,q_i)}{KT}\Big)~dp_idq_i\\
&=& \int_{{\bf R}^2}\exp\Big(-\frac{\sqrt{\lambda_i(\chi)}}{2KT}\big(p_i^2+q_i^2\big)\Big)~dp_idq_i\\
&=& \frac{2K\pi}{\sqrt{\lambda_i(\chi)}}T,
\end{eqnarray*}
and hence the internal energy is computed as 
$$
U_{\chi,i}(T)=KT. 
$$
From this, it follows that the internal energy of the solid (per the unit cell) in the classical setting is given by 
$$
U(T)=\int_{\widehat{L}}\sum_{i=1}^{3n}U_{\chi,i}(T)~d\chi=3nKT,
$$
which is nothing but the law of Dulong-Petit.

\section{Asymptotics of $\varphi(\lambda)$ at $\lambda=0$}\label{section7}

Our task is now to establish asymptotics of $U_1(T)$ and $C(T)$ as $T$ goes to zero. 
To this end, we need to study the asymptotics of $\varphi(\lambda)$ as $\lambda\downarrow 0$.

If we put 
$$
\varphi_{\chi}(\lambda)=\#\{i; \lambda_i(\chi)\leq \lambda\},
$$
then, due to (\ref{trace}), 
\begin{eqnarray*}
\varphi(\lambda)=\int_{\widehat{L}}\varphi_{\chi}(\lambda)~d\chi,
\end{eqnarray*}
which, as is easily checked, is equal to 
\begin{eqnarray*}
&&{\rm vol}\big(\{\chi\in \widehat{L};~\varphi_{\chi}(\lambda)=1\}\big)
+2\,{\rm vol}\big(\{\chi\in \widehat{L};~\varphi_{\chi}(\lambda)=2\}\big)\\
&& \quad \quad +3\,{\rm vol}\big(\{\chi\in \widehat{L};~\varphi_{\chi}(\lambda)=3\}\big)\\
&=& {\rm vol}\big(\{\chi\in \widehat{L};~\lambda_1(\chi)\leq \lambda<\lambda_2(\chi)\}\big)\\
&& \quad \quad +2\,{\rm vol}\big(\{\chi\in \widehat{L};~\lambda_2(\chi)\leq \lambda<\lambda_3(\chi)\}\big)\\
&& \quad \quad +3\,{\rm vol}\big(\{\chi\in \widehat{L};~\lambda_3(\chi)\leq \lambda\}\big)\\
&=&\sum_{\alpha=1}^3{\rm vol}\big(\{\chi;~\lambda_{\alpha}(\chi)\leq \lambda\}\big)
\end{eqnarray*}
for sufficiently small $\lambda>0$. 
Hence it suffices to establish the asymptotics of 
$$
{\rm vol}\big(\{\chi;~\lambda_{\alpha}(\chi)\leq \lambda\}\big) \quad (\alpha=1,2,3).
$$


We will use the notations introduced in the previous section and start with the operator $D$.
Note that $D$ has a ``finite radius of interaction",
that is, its kernel function $d(x,y)$ vanishes if $|\varPhi(x)-\varPhi(y)|> R$, where $R>0$
is a constant (which can be taken to be $\max_{e\in E}\{|\varPhi(te)-\varPhi(oe)|\}$).
In particular, $D$ can be naturally extended to all ${\bf C}^3$-valued functions on $V$.
This allows to decompose $D$ into direct integral of the Bloch components $D_\chi$
explicitly by taking $D_\chi$ to be the restriction of $D$ on $S_\chi$.

Let us also explicitly calculate the operator 
$$
D^0_{\chi}:C(V_0,{\bf C}^3)\longrightarrow C(V_0,{\bf C}^3).
$$ 
We claim that
\begin{eqnarray*}
D^0_{\chi}{\bf f}(x)&=&\frac{1}{m(x)}\sum_{e\in E_{0,x}}A(e)
\big\{
e^{2\pi \sqrt{-1} \chi\cdot v(e)}{\bf f}(te)-{\bf f}(oe)
\big\}. 
\end{eqnarray*}
Indeed, let us first move to the Bloch space $S_\chi$ via the correspondence 
\begin{eqnarray*}
{\bf f}\in C(V_0,{\bf C}^3) \mapsto s(x)=\exp
\big(2\pi \sqrt{-1} \chi\cdot\varPhi(x)
\big){\bf f}(\pi(x)),
\end{eqnarray*}
and take
\begin{eqnarray*}
(Ds)(x)&=&\frac{1}{m(x)}\sum_{e\in E_x}A(e)\big[e^{
2\pi \sqrt{-1} \chi\cdot\varPhi(te)}
{\bf f}(\pi(te))\\
&& \quad \quad\quad \quad\quad \quad\quad \quad -e^{
2\pi \sqrt{-1} \chi\cdot\varPhi(oe)
}{\bf f}(\pi(oe))
\big]\\
&=& e^{2\pi\sqrt{-1}\chi\cdot \varPhi(x)}\\
&& \times \frac{1}{m(x)}\sum_{e\in E_x}A(e)\big[e^{2\pi\sqrt{-1}\chi\cdot(\varPhi(te)-\varPhi(oe))}
{\bf f}(\pi(te))-{\bf f}(\pi(oe))
\big]\\
&=& e^{2\pi\sqrt{-1}\chi\cdot \varPhi(x)} \frac{1}{m(x)}\sum_{e\in E_{0\pi(x)}}A(e)\big[e^{2\pi\sqrt{-1}\chi\cdot v(e)}
{\bf f}(te)-{\bf f}(oe)
\big].
\end{eqnarray*}
Dividing by $e^{2\pi\sqrt{-1}\chi\cdot \varPhi(x)}$, we get back to $C(V_0,{\bf C}^3)$
and obtain the desired expression for $D_\chi^0$.

Let us take $\chi\in {\bf R}^3$. 
Since $D^0_{t\chi}$ is a one-parameter family of symmetric matrices depending 
analytically on $t\in {\bf R}$ (where $|t|$ is small enough so that $t\chi \in U(0)$), 
one can apply the following result (see \cite{kato}).

\begin{theorem}\label{analytic} Let $A(t)$ be an analytic family of symmetric matrices of degree $n$. Then there exist analytic functions $\lambda_1(t),\ldots,\lambda_n(t)$ $($called analytic branches of eigenvalues$)$ and orthonormal basis $u_1(t),\ldots,u_n(t)$ depending analytically on $t$ such that 
$$
A(t)u_i(t)=\lambda_i(t)u_i(t).
$$
\end{theorem}

The acoustic branches $\lambda_{\alpha}(\chi)$ are not necessarily analytic in $\chi$ because 
they are possibly bifurcated, not only at $\chi={\bf 1}$, but also at other $\chi$. The theorem above, 
however, says that one can find $\epsilon>0$ such that $\lambda_{\alpha}(t\chi)$ ($\alpha=1,2,3$) 
are real analytic in $t\in [0,\epsilon)$. To see this, take continuous branches 
$\mu_1(t),\mu_2(t),\mu_3(t)$ of eigenvalues  of $D_{t\chi}^0$ with $\mu_i(0)=0$, and let 
$I_{ij}=\{t\in {\bf R};~\mu_i(t)=\mu_j(t)\}$ ~$(i\neq j)$. Then $I_{ij}$ is either discrete or ${\bf R}$. Put 
$$
\epsilon=\min \{|t|;~t\neq 0,~t\in I_{ij}\neq {\bf R},~ i\neq j\}. 
$$
If $0< t<\epsilon$, then $\mu_i(t)\neq \mu_j(t)$, or $\mu_i\equiv \mu_j$, so that 
$\mu_j(t)$ are analytic on $[0,\epsilon)$ for all $j=1,2,3,$ and $\lambda_{\alpha}(t\chi)$ coincides with one of $\mu_1(t),\mu_2(t),\mu_3(t)$ on $[0,\epsilon)$. 

Since $\mu_j(t)\ge 0$ for all $t\in{\bf R}$, $j=1,2,3,$ we conclude that
$$
s_{\alpha}(\chi)^2:=\lim_{t\downarrow 0}\frac{1}{t^2}\lambda_{\alpha}(t\chi) \quad (s_{\alpha}(\chi)\geq 0)
$$
exists. The quantity 
$$
(2\pi\|\chi\|)^{-1}s_{\alpha}(\chi)
$$ 
is said to be the {\it acoustic phase velocity} for the direction $\chi$. It turns out to be 
the phase velocity of elastic waves in the uniform elastic body corresponding to the crystal lattice 
(see Section \ref{remark}).  

Recall that we assume $\lambda_1(\chi)\le \lambda_2(\chi)\le \lambda_3(\chi)$.
From the definition of $s_{\alpha}(\chi)$ we deduce that $s_1(\chi)\leq s_2(\chi)\leq s_3(\chi)$, 
and it follows from Theorem \ref{theorem} below that $s_{\alpha}(\chi)$ is continuous in $\chi$. 
 We also have  
$$
s_{\alpha}(t\chi)=ts_{\alpha}(\chi) \quad (t\geq 0).
$$

A perturbation argument leads to 

\begin{theorem}\label{theorem} $1)$~ $s_{\alpha}(\chi)^2$ ~$(\alpha=1,2,3)$ are eigenvalues of the symmetric matrix
$$
A_{\chi}:=\frac{2\pi^2}{m(V_0)}\sum_{e\in E_0}(\chi\cdot v(e))^2A(e).
$$ 
In particular, $s_{\alpha}(\chi)>0$ for $\chi\neq 0$, and 
$$
s_1(\chi)^2+s_2(\chi)^2+s_3(\chi)^2={\rm tr}~A_{\chi}=\frac{2\pi^2}{m(V_0)}\sum_{e\in E_0}(\chi\cdot v(e))^2{\rm tr}~A(e).
$$

$2)$~ $s_{\alpha}(\chi)$ are piecewise analytic in $\chi\in {\bf R}^3$ in the sense that there exists a proper real-analytic subset $\mathcal{S}$ in ${\bf R}^3$ such that $s_{\alpha}$ is analytic on ${\bf R}^3\backslash \mathcal{S}$.

$3)$~ The Lebesgue measure of the set $\{\chi\in {\bf R}^3;~s_{\alpha}(\chi)=1\}$ is zero. 
\end{theorem}

{\it Proof}\quad We put $D_t=D^0_{t\chi}$ and $\lambda_{\alpha}(t)=\lambda_{\alpha}(t\chi)$. In view of Theorem \ref{analytic}, one can find an orthonormal system ${\bf f}_{1,t},{\bf f}_{2,t},{\bf f}_{3,t}\in C(V_0,{\bf C}^3)$ depending analytically on $t\geq 0$ for small $t$, so that 
\begin{equation}\label{pert}
-D_t{\bf f}_{\alpha,t}=\lambda_{\alpha}(t){\bf f}_{\alpha,t}\quad (\alpha=1,2,3).
\end{equation}
Recall that ${\bf f}_{\alpha,0}$ is a constant function. If we define $e_{\alpha}(\chi)$ by 
$$
e_{\alpha}(\chi)=m(V_0)^{1/2}{\bf f}_{\alpha,0}(x),
$$
then we observe that $\{e_1(\chi),e_2(\chi),e_3(\chi)\}$ forms an orthonormal basis of ${\bf C}^3$. 

Differentiating both sides of (\ref{pert}) with respect to $t$ at $t=0$, we obtain 
\begin{eqnarray*}
&&-\frac{1}{m(x)}\sum_{e\in E_{0x}}A(e)\big[
2\pi\sqrt{-1}(\chi\cdot v(e)){\bf f}_{\alpha,0}(te)+\dot{\bf f}_{\alpha,0}(te)
-\dot{\bf f}_{\alpha,0}(oe)
\big]\\
&=&\dot{\lambda}_{\alpha}(0){\bf f}_{\alpha,0}(x)+\lambda_{\alpha}(0)\dot{\bf f}_{\alpha,0}(x).
\end{eqnarray*}
Since ${\bf f}_{\alpha,0}$ is constant, and $\dot{\lambda}_{\alpha}(0)=\lambda_{\alpha}(0)=0$, by using (\ref{inv}), we have 
$$
\frac{1}{m(x)}\sum_{e\in E_{0x}}A(e)\big(\dot{\bf f}_{\alpha,0}(te)
-\dot{\bf f}_{\alpha,0}(oe)\big)=0
$$ 
so that $\dot{\bf f}_{\alpha,0}$ is constant. 

Differentiating \eqref{pert} twice, we obtain 
\begin{eqnarray*}
&&-\frac{1}{m(x)}\sum_{e\in E_{0x}}A(e)\big[
-4\pi^2(\chi\cdot v(e))^2{\bf f}_{\alpha,0}(te)\\
&& \quad \quad \quad \quad +4\pi\sqrt{-1} \chi\cdot v(e)\dot{\bf f}_{\alpha,0}(te)+\ddot{\bf f}_{\alpha,0}(te)-\ddot{\bf f}_{\alpha,0}(oe)
\big]\\
&=&\ddot{\lambda}_{\alpha}(0){\bf f}_{\alpha,0}(x).
\end{eqnarray*}
Using the fact that $\dot{\bf f}_{\alpha,0}$ is constant, and again (\ref{inv}), we have 
\begin{eqnarray*}
&&-\frac{1}{m(x)}\sum_{e\in E_{0x}}A(e)\big[
-4\pi^2(\chi\cdot v(e))^2{\bf f}_{\alpha,0}(te)+\ddot{\bf f}_{\alpha,0}(te)-\ddot{\bf f}_{\alpha,0}(oe)
\big]\\
&=&\ddot{\lambda}_{\alpha}(0){\bf f}_{\alpha,0}(x).
\end{eqnarray*}
Here taking the inner product with ${\bf f}_{\beta,0}$, and noting 
$$
\langle D\ddot{\bf f}_{\alpha,0},{\bf f}_{\beta,0}\rangle=\langle 
\ddot{\bf f}_{\alpha,0},D{\bf f}_{\beta,0}
\rangle=0,
$$
we find 
\begin{eqnarray*}
&&4\pi^2\sum_{e\in E_0}\big(\chi\cdot v(e)\big)^2A(e){\bf f}_{\alpha,0}(te)\cdot 
\overline{{\bf f}_{\beta,0}(oe)}\\
&=&\ddot{\lambda}_{\alpha}(0)
\sum_{x\in V_0}{\bf f}_{\alpha,0}(x)\cdot \overline{{\bf f}_{\beta,0}(x)}m(x)\\
&=& \ddot{\lambda}_{\alpha}(0)\delta_{\alpha\beta}=2s_{\alpha}(\chi)^2\delta_{\alpha\beta},
\end{eqnarray*}
or, equivalently,
$$
A_{\chi}e_{\alpha}(\chi)\cdot \overline{e_{\beta}(\chi)}=s_{\alpha}(\chi)^2\delta_{\alpha\beta},
$$
and hence $s_{\alpha}(\chi)^2$ is an eigenvalue of $A_{\chi}$ with the eigenvector 
$e_{\alpha}(\chi)$. Since we assumed that $A(e)$ is strictly positive for every $e$,
we obtain also that $s_\alpha(\chi)>0$, $\alpha=1,2,3$. 

To show 2), let 
$
{\bf D}={\bf D}(\chi)
$ 
be the discriminant of the cubic polynomial 
$$
\det(zI-A_{\chi})=z^3+b(\chi)z^2+c(\chi)z+d(\chi).
$$
If ${\bf D}\not\equiv 0$, then $s_{\alpha}(\chi)$ $~(\alpha=1,2,3)$ are distinct and analytic on ${\bf R}^3\backslash \mathcal{S}_0$ where 
$$
\mathcal{S}_0=\{\chi\in {\bf R}^3;~{\bf D}(\chi)=0\}. 
$$

In the case ${\bf D}\equiv 0$, define the polynomial ${\bf D}_1(\chi)$ by 
$$
{\bf D}_1(\chi)=2b(\chi)^3-9b(\chi)c(\chi)+27d(\chi).
$$
Note that the equation $\det(zI-A_{\chi})=0$ has a root of multiplicity one (hence another root has multiplicity two) if and only if ${\bf D}_{1}(\chi)\neq 0$. If ${\bf D}_1\equiv 0$, then 
$$
s_1(\chi)^2=s_2(\chi)^2=s_3(\chi)^2=-\frac{1}{3}b(\chi)=
\frac{1}{3}\frac{2\pi^2}{m(V_0)}\sum_{e\in E_0}(\chi\cdot v(e))^2{\rm tr}~A(e),
$$
hence $s_{\alpha}$ is analytic on ${\bf R}^3\backslash \{0\}$. In the case ${\bf D}_1\not\equiv 0$, 
define  a  proper analytic subset $\mathcal{S}_1$ by
$$
\mathcal{S}_1=\{\chi \in {\bf R}^3;~{\bf D}_1(\chi)=0\}. 
$$
One can easily check that $b(\chi)^2-3c(\chi)\neq 0$ on ${\bf R}^3\backslash\{0\}$ 
and the roots of $\det(zI-A_{\chi})=0$ are 
\begin{eqnarray*}
&& \frac{9d(\chi)-b(\chi)c(\chi)}{2\big(b(\chi)^2-3c(\chi)\big)}~~\quad \quad \quad \quad\quad 
(\text{multiplicity two}),\\
&& \frac{-b(\chi)^3+4b(\chi)c(\chi)-9d(\chi)}{b(\chi)^2-3c(\chi)}\quad (\text{multiplicity one}),
\end{eqnarray*}
so that $s_{\alpha}$ is analytic on ${\bf R}^3\backslash \mathcal{S}_1$. 

The assertion 3) easily follows from 2), if we take into account the explicit form of $A_\chi$
and strict positivity of $A(e)$. 

This completes the proof of the theorem. \quad $\square$

\medskip

Think of $J_{L}$ as the flat torus with the flat metric induced from the Euclidean metric on ${\bf R}^3$. The normalized Haar measure on $\widehat{L}$ is identified with 
$$
{\rm vol}(J_{L})^{-1}d\chi,
$$
where $d\chi$, in turn, denotes the Lebesgue measure on the Euclidean space ${\bf R}^3$. Note that ${\bf V}={\rm vol}(J_{L})^{-1}$ is the volume of a unit cell. 

Put $$
A_{\alpha}(\lambda)=\{\chi\in {\bf R}^3;~\lambda_{\alpha}(\chi)\leq \lambda\},
$$
and denote by $1_{A}$ the characteristic function for a subset $A$ in ${\bf R}^3$. Then  
\begin{eqnarray*}
{\rm vol}(\{\chi \in \widehat{L};~\lambda_{\alpha}(\chi)\leq \lambda\})
&=&{\bf V}\int_{A_{\alpha}(\lambda)}~d\chi={\bf V}\int_{{\bf R}^3} 1_{A_{\alpha}(\lambda)}(\chi)~d\chi\\
&=&{\bf V}\lambda^{3/2}\int_{{\bf R}^3} 1_{A_{\alpha}(\lambda)}(\sqrt{\lambda}\chi)~d\chi.
\end{eqnarray*}
Since 
\begin{eqnarray*}
1_{A_{\alpha}(\lambda)}(\sqrt{\lambda}\chi)=\begin{cases}
1 & \text{if}~\lambda^{-1}\lambda_{\alpha}(\sqrt{\lambda}\chi)\leq 1\\
0 & \text{otherwise},
\end{cases}
\end{eqnarray*}
 we observe, due to Theorem \ref{theorem}, that 
$$
\lim_{\lambda\downarrow 0}1_{A_{\alpha}(\lambda)}(\sqrt{\lambda}\chi)=1_{A_{\alpha}}(\chi) \quad a. e.
$$
where
$$
A_{\alpha}=\{\chi\in {\bf R}^3;~s_{\alpha}(\chi)\leq 1\}. 
$$
From this, it follows that
\begin{eqnarray*}
&&\lim_{\lambda\downarrow 0}{\rm vol}(\{\chi \in \widehat{L};~\lambda_{\alpha}(\chi)\leq \lambda\}\lambda^{-3/2}\\
&=&{\bf V}\int_{{\bf R}^3}1_{A_{\alpha}}(\chi)~d\chi.
\end{eqnarray*}
Using the polar coordinates $(r,\Omega)\in {\bf R}_{+}\times S^2$, we obtain
\begin{eqnarray*}
\int_{{\bf R}^3}1_{A_{\alpha}}(\chi)~d\chi=
\int_{\{(r,\Omega);~s_{\alpha}(\Omega)r\leq 1\}}~r^2drd\Omega=
\frac{1}{3}
\int_{S^2}\frac{1}{s_{\alpha}(\Omega)^3}~d\Omega.
\end{eqnarray*}

In conclusion, we have 

\begin{theorem} 
\begin{eqnarray*}
\lim_{\lambda\downarrow 0}\varphi(\lambda)\lambda^{-3/2}=
\frac{1}{3}{\bf V}\int_{S^2}\sum_{\alpha=1}^3\frac{1}{s_{\alpha}(\Omega)^3}~d\Omega.
\end{eqnarray*}
\end{theorem}

\medskip

\noindent{\bf Remark.}\quad The asymptotic formula for an acoustic branch 
$$
\sqrt{\lambda_{\alpha}(r\chi)}\sim rs_{\alpha}(\chi) \quad (r\downarrow 0)
$$
is said to be the {\it linear dispersion law}. 

\section{Rigorous derivation of the $T^3$-law}
We are now ready to prove the $T^3$-law. 
Put 
\begin{equation}\label{c0}
c_0=\frac{1}{3}{\bf V}\int_{S^2}\sum_{\alpha=1}^3\frac{1}{s_{\alpha}(\Omega)^3}~d\Omega. 
\end{equation}
We have shown in the previous section 
$$
\varphi(\lambda)\sim c_0\lambda^{3/2} \quad \text{as}~\lambda\downarrow 0.
$$
Making the change of variables $x=$ $\hslash\sqrt{\lambda}/KT$ in the integral 
$$
U_1(T)=\int_0^{\infty}\frac{\hslash\sqrt{\lambda}}{e^{\hslash\sqrt{\lambda}/KT}-1}~d\varphi(\lambda),
$$
we obtain
\begin{eqnarray*}
U_1(T)=\hslash^{-3}K^4T^4
\int_0^{\infty}\frac{x}{e^x-1}(w_T)^{-3}d\varphi\big((w_Tx)^2\big),
\end{eqnarray*}
where $w_T=KT/\hslash$. Using $\varphi(\lambda)\sim c_0\lambda^{3/2}$, we have 
$$
\lim_{T\downarrow 0}(w_T)^{-3}\varphi\big((w_Tx)^2\big)=c_0x^3,
$$
and hence 
\begin{eqnarray*}
\lim_{T\downarrow 0}U_1(T)T^{-4}&=&3\hslash^{-3}K^4c_0\int_0^{\infty}\frac{x^3}{e^x-1}~dx\\
&=&\frac{1}{5}\pi^4c_0\hslash^{-3}K^4.
\end{eqnarray*}
Here we have used the well-known identity 
$$
\int_0^{\infty}\frac{x^3}{e^x-1}~dx=3!\zeta(4)=3!\frac{\pi^4}{90}=\frac{\pi^4}{15}.
$$
Similarly we have 
\begin{eqnarray*}
C(T)=\hslash^{-3}K^4T^3
\int_0^{\infty}\frac{x^2e^x}{(e^x-1)^2}(w_T)^{-3}d\varphi\big((w_Tx)^2\big),
\end{eqnarray*}
hence 
$$
\lim_{T\downarrow 0}C(T)T^{-3}=3\hslash^{-3}K^4c_0\int_0^{\infty}\frac{x^4e^x}{(e^x-1)^2}dx= \frac{4}{5}\pi^4c_0\hslash^{-3}K^4c_0
$$
since 
$$
\int_0^{\infty}\frac{x^4e^x}{(e^x-1)^2}dx=4!\zeta(4)=\frac{4}{15}\pi^4.
$$

Summarizing our computation, we have 

\begin{theorem}
 As $T\downarrow 0$,
\begin{eqnarray*}
&&U_1(T)\sim \frac{1}{5}\pi^4c_0\hslash^{-3}K^4T^4,\\
&&C(T)\sim \frac{4}{5}\pi^4c_0\hslash^{-3}K^4T^3.
\end{eqnarray*}
\end{theorem}

\section{Isotropic case}\label{isotropic1}
In order to compare our result with Debye's asymptotic formula, we suppose that the continuum limit of the crystal lattice is {\it isotropic}, that is, the symmetrized elastic constant tensor 
$$
A_{ij}=\frac{1}{2{\bf V}}\sum_{e\in E_{0}}v(e)_iv(e)_j A(e)
$$
satisfies 
$$
\sum_{i,j=1}^3(A_{ij})_{\alpha\beta}\chi_i\chi_j
=(a+b)\chi_{\alpha}\chi_{\beta}+b\delta_{\alpha\beta}\|\chi\|^2,
$$
where $\chi=(\chi_1,\chi_2,\chi_3)$, and $a,b$ are positive constants which are  what we call 
{\it Lame's constants} in the theory of elastic bodies. Recalling $A_\chi$ which was introduced in Theorem \ref{theorem}, we obtain 
$$
A_{\chi}=\sum_{i,j=1}^34\pi^2\rho^{-1}\chi_i\chi_jA_{ij}\quad \quad (\rho=m(V_0)/{\bf V}).
$$
The eigenvalues of $A_{\chi}$ are
\begin{eqnarray*}
&& 4\pi^2(a+2b)\rho^{-1}\|\chi\|^2 \quad (\text{multiplicity one}),\\
&& 4\pi^2 b\rho^{-1}\|\chi\|^2 \quad \quad \quad \quad (\text{multiplicity two}),
\end{eqnarray*}
since 
$$
\sum_{\beta=1}^3(A_{\chi})_{\alpha\beta}\chi_{\beta}=4\pi^2(a+2b)\rho^{-1}\|\chi\|^2\chi_{\alpha},
$$
and if ${\bf x}\cdot \chi=0$ for ${\bf x}=(x_1,x_2,x_3)$, then 
$$
\sum_{\beta=1}^3(A_{\chi})_{\alpha\beta}x_{\beta}=4\pi^2b\rho^{-1}\|\chi\|^2x_{\alpha}.
$$
Therefore  
\begin{equation}\label{isotropic2}
s_1(\chi)=s_2(\chi)=2\pi\sqrt{\frac{b}{\rho}}\|\chi\|,~~s_3(\chi)=2\pi\sqrt{\frac{a+2b}{\rho}}\|\chi\|.
\end{equation}
According to the theory of elastic waves, we say that $c_l=\displaystyle\sqrt{\frac{a+2b}{\rho}}$ is the {\it longitudinal phase velocity}, and $c_t=\displaystyle\sqrt{\frac{b}{\rho}}$ is the {\it transverse phase velocity}. 
In terms of these phase velocities, we have, by substituting (\ref{isotropic2})  for (\ref{c0})
$$
c_0=\frac{\bf V}{6\pi^2}\Big(\frac{1}{c_l^3}+\frac{2}{c_t^3}\Big).
$$ 
We thus recover Debye's result for the isotropic case. 

\section{Final remarks}\label{remark}

In general, the ``plane wave" 
$$
{\bf f}(t,{\bf x})=\exp\left[\sqrt{-1}\big(2\pi {\bf x}\cdot\chi\pm ts_{\alpha}(\chi)\big)\right]
e_{\alpha}(\chi) 
$$
in a uniform elastic body is a solution of the elastic wave equation
$$
\rho\frac{\partial^2 {\bf f}}{\partial t^2}=\sum_{i,j=1}^3 A_{ij}\frac{\partial^2{\bf f}}{\partial x_i\partial x_j}. 
$$
To get real solutions we can take real or imaginary parts
(or their real linear combinations).  For example,  
$$
{\bf f}(t,{\bf x})=\cos\big(2\pi {\bf x}\cdot\chi\pm ts_{\alpha}(\chi)\big)e_{\alpha}(\chi) 
$$
is a real plane wave which solves the same equation.

It is interesting to point out that the ``discrete plane wave" 
$$
{\bf f}_{\chi}(t,x)=\exp\left[\sqrt{-1}\big(2\pi \chi\cdot \varPhi(x)\pm t\sqrt{\lambda_{\alpha}(\chi)}\big)\right]
{\bf g}_{\chi}(\pi(x))
$$
in the crystal lattice is a solution of the equation
$$
\frac{d^2{\bf f}}{dt^2}=D{\bf f}
$$
provided that $D^0_{\chi}{\bf g}_{\chi}=\lambda_{\alpha}(\chi){\bf g}_{\chi}$. 
(The real discrete plane waves can be again obtained by taking linear combinations of the real and imaginary parts, which do not, however, have the cosine or sine forms because 
the operator $D_\chi^0$ and the eigenvectors ${\bf g}_{\chi}$ are not generally real.)

We easily observe that, for an acoustic branch $\lambda_{\alpha}(\chi)$ with 
$\lim_{\delta\downarrow 0}{\bf g}_{\delta\chi}=e_{\alpha}(\chi)$, we have  
\begin{eqnarray*}
\lim_{\delta\downarrow 0}{\bf f}_{\delta\chi}(t_{\delta},x_{\delta})=
\exp\left[\sqrt{-1}\big(2\pi {\bf x}\cdot\chi\pm ts_{\alpha}(\chi)\big)\right]
e_{\alpha}(\chi), 
\end{eqnarray*}
where the sequences $\{t_{\delta}\}$ and $\{x_{\delta}\}$ are supposed to satisfy 
$\lim_{\delta\downarrow 0} \delta t_{\delta}=t$ and 
$\lim_{\delta\downarrow 0}\delta \varPhi(x_{\delta})={\bf x}$. This justifies 
the statement that the lattice vibrations approch elastic waves in the continuum limit. 

Finally, we shall check, using the von Neumann trace again, that the function $c_0\lambda^{3/2}$ coincides with the integrated density of states for the elastic waves (thus justifying Debye's observation in his continuum theory). For this, we put 
$$
\mathcal{D}=\rho^{-1}\sum_{i,j=1}^3A_{ij}\frac{\partial^2}{\partial x_i\partial x_j},
$$
and let $K(t,{\bf x},{\bf y})$ be the kernel function of the operator $e^{t\mathcal{D}}$, namely it is the fundamental solution of the parabolic equation 
$$
\frac{\partial {\bf f}}{\partial t}=\mathcal{D}{\bf f}.
$$
Then the $L$-trace of $e^{t\mathcal{D}}$ is given as 
$$
{\rm tr}_L e^{t\mathcal{D}}=\int_{P}{\rm tr}~K(t,{\bf x},{\bf x})~d{\bf x},
$$
where $P$ is a unit cell. We can easily show, by using the Fourier transformation, 
$$
K(t,{\bf x},{\bf y})=(2\pi)^{-3}\int_{{\bf R}^3}e^{-tA(\chi)+\sqrt{-1}({\bf x}-{\bf y})\cdot \chi}~d\chi,
$$
where
$$
A(\chi)=\rho^{-1}\sum_{i,j=1}^3\chi_i\chi_jA_{ij}=(4\pi^2)^{-1}A_{\chi}.
$$
Therefore
\begin{eqnarray*}
&&{\rm tr}_L e^{t\mathcal{D}}\\
&=&{\bf V}~{\rm tr}~K(t,{\bf x},{\bf x})=
(2\pi)^{-3}{\bf V}\int_{S^2}d\Omega \int_0^{\infty} r^2~{\rm tr}~e^{-r^2tA(\Omega)}~dr\\
&=& (2\pi)^{-3}{\bf V}\int_{S^2}d\Omega \int_0^{\infty} r^2 \sum_{\alpha=1}^3
\exp\Big(-\frac{tr^2}{4\pi^2}s_{\alpha}(\Omega)^2\Big)~dr.
\end{eqnarray*}
Using the equality 
$$
\int_{0}^{\infty}r^2 e^{-ar^2}~dr=\frac{\sqrt{\pi}}{4}a^{-3/2}, \quad a>0,
$$
we get 
$$
{\rm tr}_L e^{t\mathcal{D}}=\frac{\sqrt{\pi}}{4}{\bf V}t^{-3/2}\int_{S^2}\sum_{\alpha=1}^3s_{\alpha}(\Omega)^{-3}~d\Omega.
$$
If we denote by $\varphi_0(\lambda)$ the integrated density of states for elastic waves, that is, $L$-trace of the projections in the spectral resolution of $\mathcal{D}$, then
$$
\int_{0}^{\infty}e^{-\lambda t}~d\varphi_0(\lambda)={\rm tr}_L e^{t\mathcal{D}}=\frac{3}{4}\sqrt{\pi}c_0t^{-3/2},
$$
so that, taking the inverse Laplace transform, we obtain
$$
\varphi_0(\lambda)=\frac{{\bf V}}{3}\lambda^{3/2}\int_{S^2}\sum_{\alpha=1}^3s_{\alpha}(\Omega)^{-3}~d\Omega=c_0\lambda^{3/2}
$$ 
as desired (recall the identity $\displaystyle\int_{0}^{\infty}e^{-\lambda t}~d(\lambda^{3/2})=\frac{3}{4}\sqrt{\pi}t^{-3/2}$).

\end{document}